\documentclass[%
reprint,
amsmath,amssymb,
amssthm, 
aps,
]{revtex4-2}

\usepackage{multirow}

\usepackage{xcolor,soul,framed}
\usepackage{bbold}
\usepackage{stfloats}
\usepackage[normalem]{ulem}
\usepackage{ifthen}
\usepackage{enumerate}
\newboolean{underlined}
\setboolean{underlined}{false} 
\DeclareRobustCommand{\ul}[1]{\ifthenelse{\boolean{underlined}}{\uline{#1}}{#1}}

\newboolean{strikeout}
\setboolean{strikeout}{false} 
\DeclareRobustCommand{\st}[1]{\ifthenelse{\boolean{strikeout}}{\sout{#1}}{#1}}

\usepackage{braket}
\usepackage{physics}
\usepackage{natbib}
\usepackage{graphics}
\usepackage{graphicx}
\usepackage{appendix}

\usepackage{amsthm}
\usepackage{thmtools}

\declaretheoremstyle[
  headfont=\normalfont\bfseries,
  notefont=\normalfont\bfseries,
  bodyfont=\normalfont\itshape,
  headpunct=,
]{myplainTheo}

\theoremstyle{myplainTheo}
\newtheorem{mytheorem}{Theorem}
\newtheorem{corollary}{Corollary}

\begin{document}
\preprint{APS/123-QED}

\title{Can optimal collective measurements outperform
individual measurements \\ for non-orthogonal QKD signals?}

\author{Isabella Cerutti}
\author{Petra F. Scudo}\email{petra.scudo@ec.europa.eu}
\affiliation{European Commission Joint Research Centre (JRC), 21027, Ispra, Italy}

\date{\today}

\begin{abstract}
 We consider how the theory of optimal quantum measurements determines the maximum information available to the receiving party of a quantum key distribution (QKD) system employing linearly independent but non-orthogonal quantum states. Such a setting is characteristic of several practical QKD protocols. Due to non-orthogonality, the receiver is not able to discriminate unambiguously between the signals. 
 To understand the fundamental limits that this imposes, the quantity of interest is the maximum mutual information between the transmitter (Alice) and the receiver, whether legitimate (Bob) or an eavesdropper (Eve).  
 
 To find the optimal measurement -- taken individually or collectively -- ,  we use a framework based on operator algebra and general results derived from singular value decomposition, achieving optimal solutions for von Neumann measurements and positive operator-valued measures (POVMs). 
The formal proof and quantitative analysis elaborated for two signals allow one to conclude that optimal von Neumann measurements are uniquely defined and provide a higher information gain compared to POVMs. Interestingly, collective measurements not only do not provide additional information gain with respect to individual ones, but also suffer from a gain reduction  in the case of POVMs.
\end{abstract}

\maketitle

\section{Introduction}
In this letter, we consider a general quantum key distribution (QKD) scheme in which the sender transmits non-orthogonal quantum bits and operates with a given choice of basis. 
The aim is to address the issue of optimal detection of $N$  qubits encoded in signal states $\ket{\Psi_i}$ with $i=1,2,\dots N$. 
The optimal detection strategy may be equivalently seen as an optimal interception strategy by an eavesdropper as well as an optimal measurement strategy at the receiver's end. We denote the receiver as Bob whether he is the intended party or an eavesdropper. 

The importance of the result is that collective attacks would be more practical when quantum compunting technology has developed beyond the current level, which is exactly the circumstances for which QKD is advocated. The topic is of general interest in quantum cryptography, including time-bin encoded QKD protocols such as coherent one way~\cite{stucki2005fast}, and differential phase shift~\cite{inoue2014differential}. 

We assume a quantum key encoding scheme in which the sender, Alice, prepares signal states represented as
 $\ket{\Psi_i}_k $ which identifies the $k$-th logical qubit of signal $i$. 
 Moreover, we assume that the signal states form a set of  non-orthogonal linearly independent quantum states in the vector space of the signals. Non-orthogonal signal states are typical of QKD protocols based on coherent pulses.

In order to optimally determine the state sent by Alice, the receiver (Bob) may choose to perform an unambiguous state discrimination POVM~\cite{peres1997quantum}, or a von Neumann projective measurement~\cite{von2018mathematical}. The first strategy is generally adopted in quantum cryptography and allows for the discrimination of two signal states with certainty, leaving however Bob with a finite margin of error, referred to as 'inconclusive answer'. The second strategy consists in abandoning certainties in favour of probabilities by performing a projective measurement. This latter strategy generally increases the mutual information gain of the process and is adopted by information theorists. 

In the next sections we address the main question of the paper: can optimal collective measurements outperform individual ones when discriminating among non-orthogonal QKD states? To answer this question,  we compare the use of \emph{individual measurements} with \emph{collective measurements}, where two (or potentially more) signal states are measured jointly. To this end, we consider both collective projective (von Neumann) measurements and POVMs. The comparison is based on the optimization of mutual information gain. Finally, we discuss the use of entanglement in optimizing collective measurements. 

The solution to this problem is inspired by the seminal work by Peres and Wootters on '\textit{Optimal detection of quantum information}'~\cite{peres1991optimal}  suggesting that  combined entangled measurements perform better than individual measurements when detecting  pairs of non-orthogonal tensor product quantum states. In their work, Peres and Wootters analyze the optimal detection of pairs of spin particles aligned along three non-orthogonal directions (along the $z$ axis and tilted by 120° in opposite directions). 
Note that even though each signal state is a separable state of the two particles, their optimal detection was found to be represented by an entangled operator on the state space of the two systems. 
In a subsequent work~\cite{wootters2006distinguishing}, Wootters found a joint yet separable (non-entangled) measurement that could attain the same optimal information gain. The problem was solved for specific signal states, leaving open the question about the generalization of their results to arbitrary pairs of product states. 
Note that, analogously with the works~\cite{peres1991optimal,wootters2006distinguishing}  the property we wish to investigate is whether \textit{collective}
and specifically \textit{entangled}, 
measurements outperform individual measurements when applied to separable, non-entangled signal states and whether this property holds also in the case of states, such as the time-bin ones, that do not rely on an additive group structure property (e.g. the spin coupling and Clebsch-Gordan decomposition for quantum angular momentum states). 

Based on the seminal works by Eldar and Forney~\cite{eldar2001quantum,eldar2003semidefinite}, we address the questions by deriving the optimal measurements using both a  von Neumann projective approach or POVM operators. The optimization exploits the properties of singular value decomposition (SVD) and in particular its derivation known as the Eckart-Young theorem~\cite{eckart1936approximation}. 
Compared to the previous works by Eldar and Forney, we focus on the minimization of the average information gain, for which we provide the demonstration of uniqueness and optimality of von Neumann measurements. Moreover, our results apply to  practical QKD systems allowing us to answer the question on whether collective measurements   can be more beneficial than the individual ones.

\section{Optimal von Neumann measurements}

\subsection{Optimizing the Information Gain}
\label{sec:opt-inf-gain}
We quantify the information gained by the measurement in terms of {\it average information gain}, defined as the difference between the Shannon entropy of Alice's initial preparation $H_{in}$ and the average final entropy after performing the measurement  $H_{fin}$, ~\cite{peres1997quantum,peres1991optimal}:
\begin{equation}
I_{av} = H_{in} - H_{fin}.
\label{eq:Infgain}
\end{equation}

For $N$ equiprobable signal states $\ket{\Psi_i}, r_i=1/N (1 \leq i \leq  N)$, the initial entropy is 
\begin{equation}
H_{in}= -\sum_i r_i \log(r_i)=\log(N), \label{eq:Hin_equip}
\end{equation}
where we consider the $\log$ base $2$.
The assumption of equiprobable signals sent by Alice is important for maximizing the entropy of the source and thus determining the maximum possible information gain of the receiver. The optimization of the information gain equally applies to the intended party or to an eavesdropper.  

On the other side, the average final entropy at Bob's (Eve's) end depends on the measurement results $\{ \mu\}$ and on the specific measurement strategy. We shall later compare the optimal information gain of projective von Neumann measurements with the one obtained by applying a POVM. 

To maximize the average information gain, the measurement strategy should aim at minimizing the final entropy $H_{fin}$.
Having found a specific result $\mu$, Bob's posterior probability for state $i$ is given by Bayes' theorem~\cite{grimmett2014probability}:
\begin{equation}
Q_{i, \mu} = \frac{P_{\mu, i} \, r_i}{q_{\mu}} \quad \forall i, \mu=1, \dots N  \label{eq:Qimu}   
\end{equation}
where $P_{\mu, i}$ is the probability of detecting state $\mu\ $ if the input state was $i $ and 
\begin{equation}
 q_{\mu}=\sum_j P_{\mu, j} \, r_j  \quad \forall \mu=1, \dots N
 \label{eq:qmu}
\end{equation}
is the total probability for result $\mu$.
 For equiprobable signals $r_i=1/N$ ($1 \leq i \leq N$), Eq.~(\ref{eq:Qimu}) becomes 
 \begin{equation}
Q_{i, \mu} = \frac{P_{\mu, i} }{\sum_j P_{\mu, j} } \quad \forall i, \mu=1, \dots N  \label{eq:Qimu-equiprob}.   
\end{equation}
After result $\mu $ was obtained, the associated entropy relative to $\mu$ is
\begin{equation}
H_{\mu}= - \sum_i Q_{i, \mu} \log(Q_{i, \mu})  \quad \forall \mu=1, \dots N
\label{eq:Hmu}
\end{equation}
while the average final entropy over all possible results is
\begin{equation}
<H_{fin}> = \sum_{\mu} q_{\mu} H_{\mu}.
\label{eq:Hfin}
\end{equation}

The information gain (\ref{eq:Infgain}) is maximized when the initial entropy is maximized while the average final entropy (\ref{eq:Hfin}) is minimized.
The maximum initial entropy is achieved when the signal states are equiprobable, as assumed in Eq.~(\ref{eq:Hin_equip}). 
The average final entropy is minimized when the terms $Q_{i, \mu}$ in Eq.~(\ref{eq:Hmu}) take either of the extreme values of the probability range, i.e. 0 or 1. 
In the specific case of orthonormal signals by Alice, this corresponds at the receiver's end to performing a von Neumann measurement using the same orthonormal basis as the transmitter.   
\subsection{Determining optimal von Neumann measurements for a given set of input states}

In order to translate the above observation into a quantitative condition on Bob's measurement, it is convenient to express both Alice's input states and Bob's measurement projectors in matrix form.

Let $\ket{\Psi_i}, i=1, 2, ..., N$ be Alice's input vector states with real elements and define $\mathbf{A}$ to be the matrix having as columns these vectors expressed with respect to a given basis in the Hilbert space $\mathcal{H_A}$:
\begin{align}
\mathbf{A}= \begin{pmatrix}
   \Psi_{11} & \Psi_{21} & ... & \Psi_{N1} \\
   \Psi_{12} & \Psi_{22} & ... & \Psi_{N2}\\
   ... & ... & ... \\
   \Psi_{1N} & \Psi_{2N} & ... & \Psi_{NN}
   \end{pmatrix}.
\label{eq:Amatrix}
\end{align} 
%

Let $\{\Pi_i=\ketbra{\Phi_i}{\Phi_i}\}_i$ be a complete set of orthonormal projectors on $\mathcal{H_A}$ corresponding to Bob's von Neumann measurement. 
Analogously to (\ref{eq:Amatrix}), let us define $\mathbf{B}$ to be the matrix having as columns the vectors $\ket{\Phi_i}, i=1, 2, ..., N$ in the same basis with $\braket{\Phi_i}{\Phi_j}=\delta_{i,j}$:
\begin{align}
\mathbf{B}= \begin{pmatrix}
   \Phi_{11} & \Phi_{21} & ... & \Phi_{N1} \\
   \Phi_{12} & \Phi_{22} & ... & \Phi_{N2}\\
   ... & ... & ... \\
   \Phi_{1N} & \Phi_{2N} & ... & \Phi_{NN}
   \end{pmatrix}.
\label{eq:Bmatrix}
\end{align} 

Correspondingly, the elements $P_{\mu, i}$ in (\ref{eq:Qimu-equiprob}), which are given by
\begin{equation}
 P_{\mu,i}=\left| \bra{\Phi_{\mu}}\ket{\Psi_i}\right|^2\end{equation}
can be collected in a \textit{probability matrix} $\mathbf{P}$:
\begin{equation}
 \mathbf{P}= (\mathbf{B^T}\,\mathbf{A})\circ(\mathbf{B^T}\,\mathbf{A})^*,  
\label{eq:matprodP}
\end{equation}
where the symbol $\circ$ indicates the Hadamard product between matrices and the symbol $^*$ denotes the entry-wise complex conjugate of the matrix 

Matrix $\mathbf{P}$ is normalized on the columns, i.e.
\begin{equation}
\sum_\mu P_{\mu, i}=1, \quad \forall i   \end{equation} since each input $i$ is mapped with a certain probability onto one of the possible $N$ measurement outcomes, so that the total probability sums to 1. This follows from the measurement completeness condition and holds both for von Neumann measurements and for POVMs. 
If also
\begin{equation}
\sum_i P_{\mu, i} = 1, \quad \forall i,   
\end{equation}
then matrix $\mathbf{P}$ is normalized on both columns and rows and thus termed \textit{doubly stochastic}. 
As we show below, the double stochasticity of $\mathbf{P}$ is a necessary and sufficient condition for the determination of a unique orthogonal measurement minimizing the average final entropy in (\ref{eq:Hfin}). 

We recall here the main properties of doubly stochastic matrices ~\cite{bhatia1997}.
\begin{mytheorem}[Birkhoff's Theorem]
The set of $n \cross n$ doubly stochastic matrices is a convex set whose extreme points are the permutation matrices. 
\end{mytheorem}
A very useful corollary to Birkhoff's theorem is the following: 
\begin{corollary}
 The maximum (respectively, minimum) of a convex (respectively, concave) real-valued function on the set of doubly stochastic $n \cross n$ matrices is attained
at a permutation matrix.   
\end{corollary}

Given the convexity of the entropy function to minimize, we may apply the above corollary to \eqref{eq:Infgain} to derive that the average final entropy is minimum when $ \mathbf{P}$ is a permutation matrix (i.e., with a single element equal to 1 for each row and column, and null otherwise). This corresponds, as observed earlier, to each row of $\mathbf{Q}$ having just one element equal to 1 and the others equal to 0, $Q_{i, \mu} \in \{0;1\}$. 

Without loss of generality (and optimality), we can restrict the class of permutation matrices to the identity matrix $\mathbb{1}$ since any other permutation matrix may be obtained by applying a permutation  $\mathbf{R}$ to the identity.

Thus, the objective is to find the coefficients of  $\mathbf{B}$ such that
\begin{equation}
\mathbf{B^T}\,\mathbf{A}\rightarrow\mathbb{1}.
\label{eq:matrprodOpt}
\end{equation}
By multiplying the left side by a permutation matrix $ \mathbf{R}$, 
\begin{equation}
\mathbf{R}\,\mathbf{B^T}\,\mathbf{A}\rightarrow\mathbf{R}\,\mathbb{1}=\mathbf{R}
\label{eq:matrprodPi}
\end{equation}
it is possible to explore all the other possible optimal solutions with swapped rows and columns, leading to the same value of minimal entropy.
Relation \eqref{eq:matrprodOpt} is realized with equality only when $\mathbf{A}$ is also orthonormal, i.e., Alice sends  orthogonal signals.
For non-orthogonal signals, the optimal matrix $\mathbf{B}$ realizing \eqref{eq:matrprodOpt} is the closest orthonormal matrix which approximates $\mathbf{A}$ in the least square sense.   

In order to solve matrix equation (\ref{eq:matrprodOpt}) for unknown $\mathbf{B}$ (matrix $\mathbf{A}$ is given by Alice's choice of signal states), we make use of the Eckart-Young~\cite{eckart1936approximation} and Mirsky~\cite{mirsky1960symmetric} theorems  of linear algebra  in their "minimal transformation to orthonormality" formulation derived by Johnson~\cite{johnson1966minimal}. We briefly recall the main statement and proof of this theorem.

\begin{mytheorem}[Johnson's Theorem]
Let $\mathbf{X}$ be a $n \cross n$ complex full-rank matrix. Then the \textit{minimal transformation to orthonormality} approximating $\mathbf{X}$ is given by $\mathbf{Z}$ such that:
\begin{eqnarray}
\text{Objective function: } &\min \, Tr\{(\mathbf{X} - \mathbf{Z})^T (\mathbf{X} - \mathbf{Z})\} \label{eq:Johnsonsth}\\
\text{Constraint: } &\mathbf{Z}^T\mathbf{Z} = \mathbb{1}\label{eq:Johnsonsconstr}.
\end{eqnarray}
where $Tr$ indicates the trace of the matrix.
\end{mytheorem}
The proof of Johnson's theorem is based on the Eckart-Young formulation of the singular value decomposition (SVD), allowing the derivation of the orthonormal matrix $\mathbf{Z}$ minimizing \eqref{eq:Johnsonsth}. 
Note that the objective function (\ref{eq:Johnsonsth}) is equivalent to
\begin{equation}
 \min \, Tr\{(\mathbf{Z}^T\mathbf{X} - \mathbb{1})^T (\mathbf{Z}^T\mathbf{X} - \mathbb{1})\}, \label{eq:Johnsonsidrel}   
\end{equation}
thus translating into the closest approximation to the identity of matrix $\mathbf{Z}^T\mathbf{X}$. In our notation, Johnson's $\mathbf{X}$ matrix corresponds to Alice's  $\mathbf{A}$, while $\mathbf{Z}$ is Bob's $\mathbf{B}$.
Matrix $\mathbf{A}$ can be decomposed into the product of matrix $\mathbf{U}$ having as columns the eigenvectors of $\mathbf{A A^T}$, the diagonal singular value matrix  $\mathbf{\Sigma}$, and matrix $\mathbf{V}$ having as columns the eigenvectors of  $\mathbf{A^T A}$:
\begin{equation}
 \mathbf{A} = \mathbf{U}\, \mathbf{\Sigma} \,\mathbf{V}^T   
\end{equation}
where $\mathbf{U} \mathbf{U}^T = \mathbb{1}$ and $\mathbf{V} \mathbf{V}^T = \mathbb{1}$. 
The orthonormal matrix $\mathbf{B}$ that best approximates $\mathbf{A}$ 
\begin{equation}
\min \, Tr(\mathbf{B}-\mathbf{A})^{T}(\mathbf{B}-\mathbf{A}),    
\end{equation}
is given by 
\begin{equation}
    \mathbf{B} = \mathbf{U} \, \mathbf{V}^T
    \label{eq:optB}
\end{equation}
which is the product of the two orthonormal matrices derived from the SVD decomposition of $\mathbf{A}$. Note that, whenever $\mathbf{A}$ admits distinct singular values, the solution defined by (\ref{eq:optB}) is unique up to column permutations and yields 
\begin{equation}
    \mathbf{B}^T \, \mathbf{A}= \mathbf{V} \,\mathbf{\Sigma} \, \mathbf{V}^T,
    \label{eq:optBAproduct}
\end{equation}
which defines a symmetric matrix (as the right-side and left-side orthonormal matrices are identical).

Being symmetric,  $\mathbf{P}$ resulting from \eqref{eq:matprodP} is a \textit{doubly stochastic} matrix. 
%
Thus, if $\mathbf{P}$ is a \textit{doubly stochastic} matrix the optimal measurement matrix $\mathbf{B}$ maximizing the average information gain (\ref{eq:Infgain}) is derived by virtue of Birkhoff's theorem by (\ref{eq:matrprodPi}) and \textit{uniquely defined} by  (\ref{eq:optB}).

Vice versa, assuming that $\mathbf{B} = \mathbf{U} \, \mathbf{V}^T$ is the optimal solution to maximizing the average information gain, substituting (\ref{eq:optBAproduct}) into (\ref{eq:matprodP}) we obtain a doubly stochastic matrix. The proof of the latter implication is a straightforward consequence of the symmetry of the matrix product (\ref{eq:optBAproduct}). 
We are able therefore to state that double stochasticity is a \textit{necessary and sufficient condition} for the determination of a unique optimal orthonormal $\mathbf{B}$ of the form $\mathbf{B} = \mathbf{U} \, \mathbf{V}^T$.

So far, the search for Bob's optimal  measurement assumed a doubly stochastic matrix $\mathbf{P}$. We wish now to explore optimality relaxing this initial condition.
By definition $\mathbf{P}$ is positive semi-definite as its elements correspond to the transition probabilities between Alice's states and Bob's measurement vectors. In probabilistic terms, $\mathbf{P}$ is a stochastic matrix describing the transitions of a Markov chain~\cite{norris}. When Alice's states are linearly independent, $\mathbf{P}$ is in fact strictly positive.  
We recall here the following set of results for strictly positive $n \cross n$ matrices derived by Sinkhorn, Marcus and Newman~\cite{marcus1965generalized}, Sinkhorn and Knopp~\cite{sinkhorn1967concerning}.
\begin{mytheorem}[Sinkhorn's Theorem]
Let $\mathbf{X}$ be a strictly positive $n \cross n$ matrix. Then to $\mathbf{X}$ there corresponds a unique doubly stochastic matrix $\mathbf{T_X}$ which can be expressed in the form: 
 \begin{equation}
    \mathbf{T_X} = \mathbf{D}_1 \, \mathbf{X} \, \mathbf{D}_2,
\label{eq:Sinkhtheor}    
\end{equation}
where $\mathbf{D}_1, \, \mathbf{D}_2$ are diagonal matrices with positive entries. $\mathbf{D}_1, \, \mathbf{D}_2$ are unique themselves up to a scalar factor.
\end{mytheorem}
The doubly stochastic matrix $\mathbf{T_X}$ is derived as a limit of the sequence of matrices generated by  normalizing alternately the rows and the columns of $\mathbf{X}$, until convergence. A sufficient condition for this scaling process to converge is provided by the following:  
\begin{corollary}[Marcus and Newman]
If $\mathbf{X}$ is strictly positive and symmetric there exists a diagonal matrix $\mathbf{D}$ with positive main diagonal entries such that 
\begin{equation}
\mathbf{T_X} = \mathbf{D} \, \mathbf{X} \, \mathbf{D}  
\end{equation}
is doubly stochastic.
\label{cor:MarcusNeuman}
\end{corollary}
Sinkhorn and Knopp later showed that the sequence of matrices generated by  normalizing alternately the rows and the columns of $\mathbf{X}$ converges to the doubly stochastic limit $\mathbf{T_X} = \mathbf{D}_1 \, \mathbf{X} \mathbf{D}_2$ if and only if $\mathbf{X}\neq 0$ and each positive entry of $\mathbf{X}$ is contained in a positive diagonal.

Matrices for which the normalizing sequence converges are termed \textit{scalable}. For such matrices, the best known scaling algorithm consists in applying a 'coordinate-descending' method known as \textit{RAS} or \textit{biproportional fitting} algorithm. 
Note that in all cases in which the algorithm converges to a doubly stochastic matrix, we return to the same initial assumptions enabling the implementation of Johnson's theorem and thus to the same optimal solution (\ref{eq:optB}). The latter consideration also provides an answer to the question raised long ago in the seminal work by Hausladen and Wootters discussing the optimality of their so called 'pretty good measurement' with respect to maximizing average information gain~\cite{hausladen1994}. Their density operator $\mathbf{\rho}$ corresponds  to 
\begin{equation}
 \rho=\mathbf{A^T}\mathbf{A}, \label{eq:prettygoodA}   
\end{equation}
and thus the pretty good measurement matrix to $\mathbf{M} = \mathbf{A} (\mathbf{A^T}\mathbf{A})^{-\frac{1}{2}}$. By substituting the SVD expression $\mathbf{A} = \mathbf{U}\, \mathbf{\Sigma} \,\mathbf{V}^T$ and taking into account the unitarity of $\mathbf{U}, \mathbf{V}$ it is straightforward to check that $\mathbf{M}=\mathbf{U} \mathbf{V^T}$ coincides with the optimal $\mathbf{B}$. 
This provides a rigorous demonstration of the optimality of their pretty good measurement by virtue of the scaling properties of the probability.

We summarize our findings in the following theorem:
\begin{mytheorem}[Optimal detection of non-orthogonal signals] \label{mytheorem}
Let $\ket{\Psi_i}, i=1, 2, ..., N$ be a set of non-orthogonal linearly independent signal states on Alice's Hilbert space $\mathcal{H_A}$ and $\ket{\Phi_{\mu}}, \mu=1, 2, ..., N$ a set of orthonormal states on Bob's $\mathcal{H_B}$, with $dim(\mathcal{H_A})=dim(\mathcal{H_B})=N$. Let $\mathbf{A}, \mathbf{B}$ be the matrices having these states as column vectors 
and $\mathbf{P} =  (\mathbf{B^T}\,\mathbf{A})\circ(\mathbf{B^T}\,\mathbf{A})^*$ be the corresponding stochastic matrix of transition probabilities. The following propositions hold:
\begin{enumerate}
    \item If $\mathbf{P}$ is \text{doubly stochastic}, or
    \item If $\mathbf{P}$ is strictly positive and symmetric, or 
    \item If $\mathbf{P}$ is \textit{scalable} to a doubly stochastic limit matrix, 
\end{enumerate}
then the optimal von Neumann measurement $\mathbf{B}$ maximizing the average information gain defined in \eqref{eq:Infgain} is uniquely determined by $\mathbf{B} = \mathbf{U} \, \mathbf{V}^T$ where $\mathbf{U}$ and $\mathbf{V}$ are unitary operators diagonalizing $\mathbf{A} \, \mathbf{A}^T$ and $\mathbf{A}^T \, \mathbf{A}$, respectively. 
\end{mytheorem}

\section{Collective vs. Individual Measurements}

In the following section, we use the results derived above to compare individual with collective von Neumann measurements. The question is motivated by considering a general QKD scenario in which Alice transmits a sequence of $K$ ($K \geq 2$) signals. Instead of measuring each signal individually, Bob may decide to store the $K$ signals in a quantum memory and perform a collective measurement. We will consider the case $K=2$ independent signals as the result can then be trivially extended to any $K > 2$.
We may think of them as states belonging to consecutive time-slots (i.e. generated by QKD systems operating in time division multiplexing). If each of the signals is represented as before by $\ket{\Psi_i}, i=1, 2, ..., N$, the possible pairs are represented by $\ket{\Psi_{i,j}}, i,j=1, 2, ..., N$ in the Hilbert space of dimension $N^2$ $\mathcal{H_A}\otimes \mathcal{H_A}$:
\begin{equation}
    \ket{\Psi_{i,j}} = \ket{\Psi_{i}} \otimes\ket{\Psi_{i}} \quad \forall i,j=1, 2, ..., N
\end{equation}
where $\otimes$ is the tensor or Kronecker product.
Let $\mathbf{A_2}$ the matrix having the vectors $\ket{\Psi_{i,j}}$ as columns, or equivalently
\begin{equation}
    \mathbf{A_2} = \mathbf{A} \otimes \mathbf{A}. \label{eq:A2}
\end{equation}

Among  Bob's joint measurements we may further distinguish between collective (or joint) uncorrelated measurements described by tensor product states or collective \textit{entangled} measurements.  For example, in~\cite{peres1991optimal} it was shown that for the so called \textit{double trine} states the optimal measurement was global and entangled. Later, Wootters~\cite{wootters2006distinguishing} determined a global yet unentangled measurements that performed equally well obtaining the same optimal mutual information gain. 
Analogouesly, in our case Bob may consider collective von Neumann measurements defined by product states  
\begin{equation}
    \ket{\Phi_{\mu,\nu}} = \ket{\Phi_{\mu}} \otimes\ket{\Phi_{\nu}} \quad \forall \mu,\nu=1, 2, ..., N
\end{equation}
or by general superpositions of such states (including entangled ones)
\begin{equation}
    \ket{\Phi_{M,N}} = \sum_{\mu,\nu} B_{\mu,\nu; M, N} \ket{\Phi_{\mu}} \otimes \ket{\Phi_{\nu}}. 
\end{equation}

The corresponding elements of the probability matrix $\mathbf{P}$ are labelled by a double index measurement result and a double index input signal:
\begin{align}
    P_{{M,N;i,j}} & = |\bra{\Phi_{M,N}} \ket{\psi_{i,j}}|^2\\ 
    & = \Big|\sum_{\mu,\nu} B_{\mu,\nu; M, N} \bra{\Phi_{\mu}}\ket{\psi_i} \bra{\Phi_{\nu}} \ket{\psi_j}\Big|^2. 
\end{align}

For product states, the matrix $\mathbf{B_2}$ having the set of orthonormal vectors $\ket{\Phi_{i,j}}=\ket{\Phi_i}\ket{\Phi_j}$ as columns can be written as 
\begin{equation}
    \mathbf{B_2} = \mathbf{B} \otimes \mathbf{B}. \label{eq:B2}
\end{equation}
Analogously, for general superposition including entangled  measurements, $\mathbf{B_2}$ corresponds to linear combinations of such products
\begin{equation}
\mathbf{B_2} = \sum_{\mu,\nu} B_{\mu,\nu} \, \mathbf{B}_{\mu} \otimes \mathbf{B}_{\nu},
\end{equation}

where the multi-index $\mu$ in matrix $\mathbf{B}_{\mu}$ defines a specific permutation of the column vectors $\ket{\Phi_i}, i=1,2, ..., N$ and the superposition coefficients $B_{\mu,\nu} \in \mathbb{C}$ are normalized in square modulus $\sum_{\mu,\nu} |B_{\mu,\nu}|^2=1$.  
 
The optimal von Neumann product measurement leads to the probability matrix $\mathbf{P_2}$ as
\begin{align}
 \mathbf{P_2} &= (\mathbf{B_2}^T \cdot \mathbf{A_2}) \circ (\mathbf{B_2}^T \cdot \mathbf{A_2})^* =  \label{eq:P2}\\
 & =  \big((\mathbf{B} \otimes \mathbf{B})^T \cdot (\mathbf{A} \otimes \mathbf{A})\big) \circ 
 \big(\mathbf{B}\otimes \mathbf{B})^T \cdot (\mathbf{A}\otimes \mathbf{A})\big)^*  \nonumber\\
 & =  \big((\mathbf{B}^T \cdot \mathbf{A})  \otimes  (\mathbf{B}^T \cdot \mathbf{A})\big) \circ 
 \big((\mathbf{B}^T \cdot \mathbf{A})  \otimes  (\mathbf{B}^T \cdot \mathbf{A})\big)^* \nonumber\\
  & =  \big((\mathbf{B}^T \cdot \mathbf{A})  \circ  (\mathbf{B}^T \cdot \mathbf{A})^*\big)\otimes 
  \big((\mathbf{B}^T \cdot \mathbf{A})  \circ  (\mathbf{B}^T \cdot \mathbf{A})^*\big) \nonumber\\
  & =  \mathbf{P} \otimes \mathbf{P}   \nonumber
\end{align}
where the mixed-product property for Kronecker and Hadamard products was used, along with Eqs.~(\ref{eq:matprodP}), (\ref{eq:A2}), (\ref{eq:B2}).
For general superposition measurements, each term in (\ref{eq:P2}) $(\mathbf{B_2}^T \cdot \mathbf{A_2})$ becomes:
\begin{equation}
\Big(\sum_{\mu,\nu} B_{\mu,\nu} \mathbf{B}_{\mu} \otimes \mathbf{B}_{\nu}\Big)^T \cdot (\mathbf{A} \otimes \mathbf{A}).    
\end{equation}
To derive the optimal superposition measurement, we use the optimization  for the optimal $\mathbf{B}$ in (\ref{eq:Johnsonsth}) 
\begin{align}
 \min \, Tr&\Big\{\Big[(\mathbf{A}\otimes \mathbf{A}) - \Big(\sum_{\mu,\nu} B_{\mu,\nu} \, \mathbf{B}_{\mu} \otimes \mathbf{B}_{\nu}\Big)\Big]^T \nonumber \\
 &\Big[(\mathbf{A}\otimes \mathbf{A}) - \Big(\sum_{\mu',\nu'} B_{\mu',\nu'} \,\mathbf{B}_{\mu'} \otimes \mathbf{B}_{\nu'}\Big)\Big]\Big\},    \end{align}
which leads to the following equivalent condition, after developing the products within the trace 
\begin{align}
   & \max \, Tr{\Big(\sum_{\mu,\nu} \, B_{\mu,\nu} \, (  \mathbf{B}_{\mu}^T \, \mathbf{A}) \otimes (\mathbf{B}_{\nu}^T \, \mathbf{A})\Big)} \nonumber\\
   & = \max \, \sum_{\mu,\nu} B_{\mu,\nu} \cdot  Tr{(\mathbf{B}_{\mu}^T\,\mathbf{A})} \, \cdot Tr{(\mathbf{B}_{\nu}^T \, \mathbf{A})} \nonumber\\
   & \leq \max \, \sum_{\mu,\nu} |B_{\mu,\nu}| \cdot  Tr{(\mathbf{B}_{\mu}^T \, \mathbf{A})} \, \cdot Tr{(\mathbf{B}_{\nu}^T\, \mathbf{A})} \nonumber \\
   & \leq \max \, \big(Tr{(\mathbf{B}_{\mu}^T \, \mathbf{A})} \, \cdot Tr{(\mathbf{B}_{\nu}^T \, \mathbf{A})}\big)\label{eq:Bent},
\end{align}
under the orthogonality condition for matrices $\mathbf{B}_{\mu},\mathbf{B}_{\nu}$. 

In the above relations we used the linearity of the trace operator, the property that $Tr(\mathbf{X}\otimes\mathbf{Y})=Tr(\mathbf{X}) \cdot Tr(\mathbf{Y})$ and the fact that the maximum of a convex combination is in one of the extreme points. Recalling the derivation in Johnson's theorem in~\cite{johnson1966minimal}, determining $\max \, Tr{(\mathbf{B}_{\mu}^T \, \mathbf{A})}$ leads to a unique optimal solution $\mathbf{B}$ up to column permutations. 

The result indicates that the matrix of the collective measurement probability $\mathbf{P_2}$ can be written as a Kronecker product of the matrices representing the individual measurement probabilities $\mathbf{P}$, for both  product and superposition measurements. Thus,  there is no information gain in performing a collective measurement -- either uncorrelated or entangled --  over two (or more) non-orthogonal signals when using a von Neumann projection. 

\section{Optimal unambiguous state discrimination}

Optimal unambiguous state discrimination (USD) was first solved for the case of two signals by Ivanovic~\cite{ivanovic1987differentiate}, Dieks~\cite{dieks1988overlap} and Peres~\cite{peres1988differentiate} and for three signals by Peres and Terno~\cite{peres1998optimal}. Partial results for $N$ non-orthogonal linearly independent signals were provided by Chefles~\cite{chefles1998unambiguous} while a complete solution in terms of semidefinite programming was found by Eldar~\cite{eldar2003semidefinite} and improved recently by Karimi~\cite{karimi2021optimal}. 


Optimal USD is based on  positive operator valued measures (POVMs). Whereas a projective or von Neumann measurement generates probabilities, a USD POVM either identifies the correct state with certainty, or generates an inconclusive answer. 
Referring to the above definition of $Q_{i, \mu}$ (\ref{eq:Qimu}), applying a USD POVM we have:
\begin{equation}
    Q_{i, \mu} = \delta_{i \mu} \quad \forall i, \mu = 1, ... N, 
\end{equation}
where $\delta_{i \mu}$ is the Kronecker delta
and thus (\ref{eq:Hmu}) becomes
\begin{equation}
    H_{\mu} = 0 \quad \forall \mu=1, ... N.
\end{equation}
Therefore the only non null term contributing to the final entropy is the one associated with the inconclusive answer $\mu=0$: 
\begin{equation}
   Q_{i, 0} = \frac{P_{0, i} \, r_i}{q_{0}} \quad \forall i =1, \dots N, 
\end{equation}
and correspondingly
\begin{align}
H_{0} &= - \sum_i Q_{i, 0} \log(Q_{i, 0})  \\  \label{eq:Hmu-POVM}
<H_{fin}> &= q_{0} H_{0}  = -\sum_i P_{0, i} \log (\frac{P_{0, i}}{\sum_j P_{0, j}}).    
\end{align}

In the following the final entropy is optimized by minimizing the probability of an inconclusive result. 
Indeed, by setting $P_{0, i} = p_{inc}, \, i= 1, ...N$,
we obtain
\begin{equation}
 <H_{fin}> = p_{inc} \log(N),   
 \label{eq:Kfin_POVM}
\end{equation}
where we used equal probabilities also for Alice's input states $\ket{\Psi_i}$.

The optimal POVM operators $\{\Pi_i\}_{i=0, 1, ...N}$ such that
\begin{equation}
    \sum_{i=0}^N \Pi_i = \mathbb{1}, \quad \quad  \sum_{i=1}^N \Pi_i \leq \mathbb{1} \label{eq:sumPOVM}
\end{equation} can be derived following the demonstration in~\cite{eldar2003semidefinite}.
Let $\ket{\tilde{\Psi}}$ be the reciprocal states associated with  Alice's states $\ket{\Psi}$, such that $\braket{\Psi_h}{\tilde{\Psi_k}} = \delta_{hk} \, (1 \leq h,k \leq N)$, i.e., the reciprocal states are orthogonal to Alice's states. Define 
\begin{equation}
\Pi_i = (1-p_{inc}) \ket{\tilde{\Psi_i}}\bra{\tilde{\Psi_i}},\label{POVMelements}    
\end{equation}
and $\mathbf{\Tilde{A}}$ to be the matrix whose columns are the reciprocal states $ \ket{\tilde{\Psi_i}}, i=1, ...N$. 
The matrix $\mathbf{\Tilde{A}}$ is orthogonal to $\mathbf{{A}}$, i.e., 
\begin{equation}
    \mathbf{\Tilde{A}}^T \cdot \mathbf{A} = \mathbb{1},
\end{equation}
showing similarity with Eq. \eqref{eq:matrprodOpt} for the von Neumann case. The solution can be found by resorting to the Moore-Penrose pseudo-inverse~\cite{penrose1955}, i.e., 
\begin{equation}
\mathbf{\Tilde{A}}^T =  (\mathbf{A}^T \, \mathbf{A})^{-1} \, \mathbf{A}^T.
\end{equation}

By applying the SVD to $\mathbf{A} = \mathbf{A \, \Sigma \,V^T}$, we find that the matrix of the reciprocal states is given by
\begin{equation}
\mathbf{\Tilde{A}} = \mathbf{U} \, \mathbf{\Sigma}^{-1} \,\mathbf{V}^{T},    
\end{equation}
i.e. its singular values are the inverse of the ones of $\mathbf{A}$ but with the same eigenvector matrices $\mathbf{U}$ and $\mathbf{V}$.

For equiprobable signals, the summation in \eqref{eq:sumPOVM} is equal to
\begin{equation}
(1-p_{inc}) \sum_{i=1}^N \ket{\tilde{\Psi_i}}\bra{\tilde{\Psi_i}}   = \mathbf{\Tilde{A}}\, \mathbf{\Tilde{A}}^T = \mathbf{U} \,\mathbf{\Sigma}^{-2}\, \mathbf{U}^{T}
\end{equation}
and 
 the probability of conclusive result (i.e., $1-p_{inc}$) is maximized when it is equal to the inverse of the maximum eigenvalue of $\sum_i \braket{\tilde{\Psi_i}}{\tilde{\Psi_i}}$, or equivalently to the minimum singular value of $\mathbf{\Sigma}^{2}$.

\section{Pairs of non-orthogonal signals}

We examine first the application of the present framework to the case of a single qubit in two equally likely non-orthogonal states.  Such states are representative of the non-orthogonal signals used in some QKD protocols~\cite{huttner1994quantum}, \cite{ekert1994}. The angle between them is identified as $\theta\in(0,\pi/2]$. When $\theta=\pi/2$, the two signals become orthogonal and thus perfectly distinguishable by a standard von Neumann measurement in the same basis. The signals can be generally expressed as\begin{align}
\Psi_{1}= \begin{pmatrix}
   \cos(\frac{\theta}{2}) \\
    \sin(\frac{\theta}{2})
   \end{pmatrix} \quad
   \Psi_{2}= \begin{pmatrix}
   ~~\cos(\frac{\theta}{2}) \\
    - \sin(\frac{\theta}{2})
   \end{pmatrix}. 
\end{align} 

By replacing $c= \cos(\frac{\theta}{2})$ and $s = \sin(\frac{\theta}{2})$, the corresponding matrix $\mathbf{A}$ is given by
\begin{align}
\mathbf{A}= \begin{pmatrix}
   \cos(\frac{\theta}{2}) &~~\cos(\frac{\theta}{2}) \\
   \sin(\frac{\theta}{2}) & -\sin(\frac{\theta}{2})
   \end{pmatrix} = \begin{pmatrix}
   c&~~c\\
   s& -s
   \end{pmatrix}
\end{align} 
and  can be SVD-decomposed into
\begin{align}
\mathbf{A}&= \mathbf{U} \, \mathbf{\Sigma} \,\mathbf{V}^T=\\ 
  &= \begin{pmatrix} 
   1 &0 \\
   0& 1
   \end{pmatrix}   
   \begin{pmatrix} 
   \sqrt{2}\,c & 0\\
   0 &  \sqrt{2} \, s 
   \end{pmatrix} 
   \begin{pmatrix} 
   1/\sqrt{2} & ~~1/\sqrt{2} \\
   1/\sqrt{2} & -1/\sqrt{2}
   \end{pmatrix}.  \nonumber
\end{align} 
Based on \eqref{eq:optB}, the optimal von Neumann measurement  is thus
\begin{align}
\mathbf{B}= \mathbf{U} \, \mathbf{V}^T= \mathbf{V}^T = \frac{1}{\sqrt{2}}
   \begin{pmatrix} 
   1 & ~~1\\
   1 & -1
   \end{pmatrix} \label{B2opt},
\end{align} 
i.e., a rotation by $\pi/4$ of Alice's basis, followed by a reflection.

The associated probability matrix for individual measurements is
\begin{align}
\mathbf{P} &= (\mathbf{V}\, \mathbf{\Sigma}\, \mathbf{V}^T)\circ (\mathbf{V}\, \mathbf{\Sigma}\, \mathbf{V}^T)^*  = \frac{1}{2} 
   \begin{pmatrix} 
   (c+s)^2 & (c-s)^2\\
   (c-s)^2 & (c+s)^2
   \end{pmatrix} \nonumber\\
   &=  \frac{1}{2}    \begin{pmatrix} 
   1+ \sin(\theta)& 1-\sin(\theta)\\
   1-\sin(\theta) & 1+\sin(\theta)
   \end{pmatrix} 
\end{align}

Note that $\mathbf{P}$ is doubly stochastic since the sum of the elements on each row is equal to 1 as is the sum of the elements on each column. Indeed, each element of $\mathbf{V}$ has square value of $1/2$, which is a sufficient condition for the doubly stochasticity of $\mathbf{P}$.
When $\theta = \pi/2$ (i.e., the two signals are orthogonal), then $\mathbf{P}= \mathbb{1}$ and reaches the maximum possible entropy.

The average information gain for von Neumann measurements computed from \eqref{eq:Infgain}-\eqref{eq:Hfin} is
\begin{equation}
{I_{av}} \!= 1 \!-\! (c+s)^2  \log_2 \frac{(c+s)^2}{2} \!-\! (c-s)^2 \log_2 \frac{(c-s)^2}{2}.
\end{equation} 

With POVM measurements, the probability of an inconclusive result is derived from the
smallest singular value of $\mathbf{\Sigma} $  (i.e., $\sqrt{2} s$), leading to $p_{inc} = 1-(\sqrt{2}\,s)^2$. The average information gain is obtained by applying \eqref{eq:Infgain} and \eqref{eq:Kfin_POVM}, leading to $I_{av} = 2 \, s^2 $.

For the case of two non-orthogonal signals, a generic $SO(2)$ rotation matrix can be written as 
\begin{align}
\mathbf{B_{SO(2)}}= \begin{pmatrix}
   \cos(\phi) & - \sin(\phi) \\
   \sin(\phi) & ~~\cos(\phi)
   \end{pmatrix},
\end{align}     
with $\phi \in (0, 2\pi]$.
In this case it is easily seen that the generic $\mathbf{B_{SO(2)}}$ matrix leads to a scalable $\mathbf{P_{SO(2)}}$ in the sense of Theorem~\ref{mytheorem}, since
\begin{align}
&\mathbf{P_{SO(2)}} =
 (\mathbf{B_{SO(2)}}^T \, \mathbf{A}) \circ (\mathbf{B_{SO(2)}}^T \, \mathbf{A})^* \\
  &=
 \begin{pmatrix} 
(~~ c \! \cdot \! \cos(\phi) \!+\! s \!\cdot \!\sin(\phi))^2 & (c\! \cdot\!\cos(\phi) \!-\!s\! \cdot\! \sin(\phi))^2\\
   (c \!\cdot\!\sin(\phi) \!- \!s \!\cdot\! \cos(\phi))^2 &  (c\! \cdot \!\sin(\phi)\!+\!s\! \cdot\!\cos(\phi))^2
   \end{pmatrix} \nonumber
\end{align}
is doubly stochastic $\forall \theta \in (0, \pi/2]$ if and only if 
\begin{equation}
    \sin^2(\phi) = \cos^2(\phi) \quad \Rightarrow \quad \sin (\phi) = \pm \cos(\phi), 
\end{equation}
leading to the same optimal $\mathbf{B}$ derived in Eq.~(\ref{B2opt}) and demonstrating the uniqueness of the optimal solution and its convergence.

\begin{figure}[htb]
    \centering
    \includegraphics[width=0.9 \columnwidth]{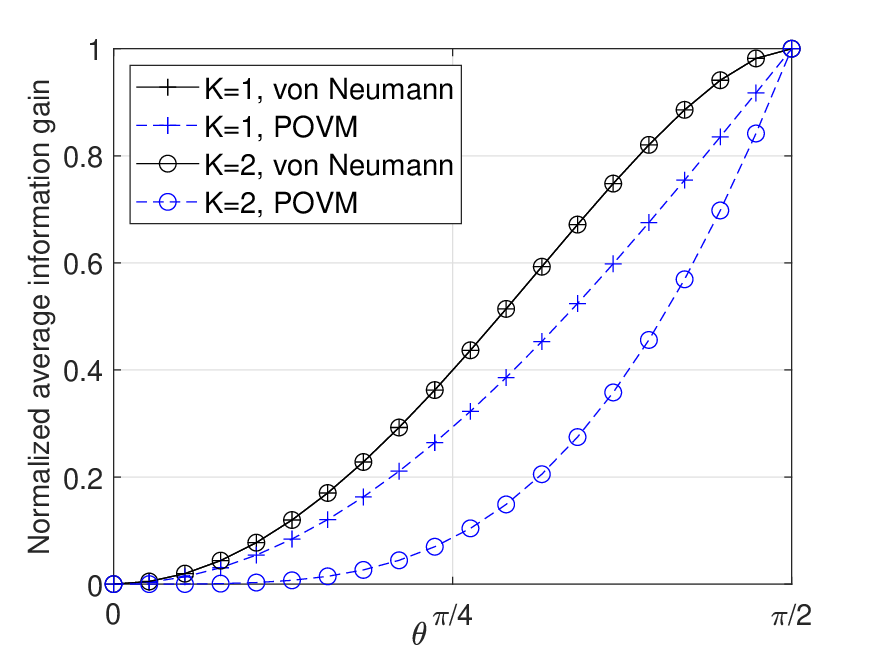}
    \caption{Avg. information gain vs. $\theta$ for individual ($K=1$) and collective ($K=2$) measures normalized to $K$}
    \label{fig:avgInfGain}
\end{figure}

\begin{figure}[htb]
    \centering
    \includegraphics[width=0.9 \columnwidth]{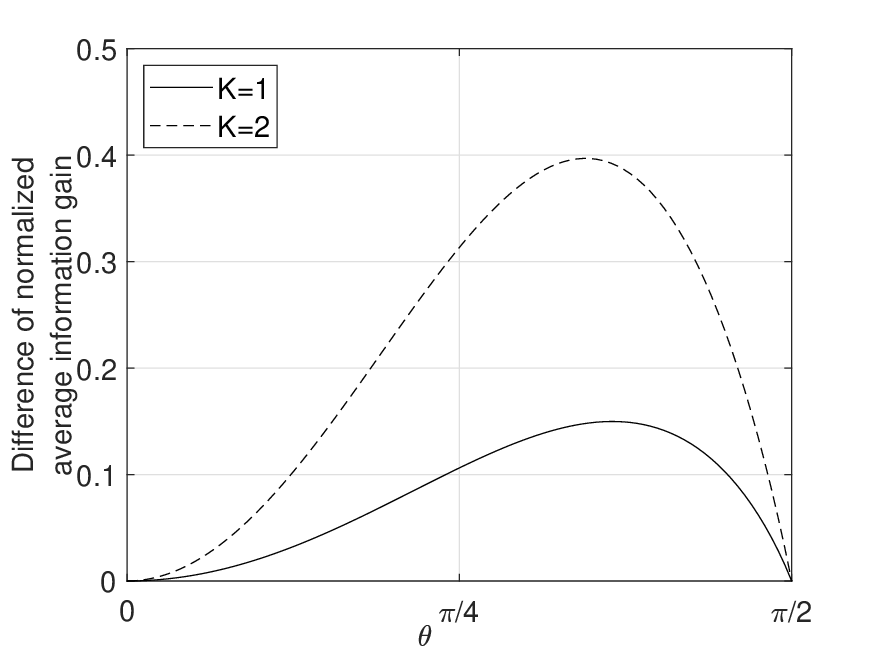}
    \caption{Difference of normalized avg. information gain (between von Neumann and POVM measurements) as a function of $\theta$, for individual ($K=1$) and collective ($K=2$) measures}
    \label{fig:diffAvgInfGainC}
\end{figure}

For collective measurements on pairs of two signals $\Psi_i \otimes \Psi_j (i,j= 1,2)$, the resulting matrix $\mathbf{A}$ is:
\begin{align}
\mathbf{A_2}&\!=\! \begin{pmatrix}
   c^2  & ~~c^2 &  ~~c^2  & ~~c^2 \\
   c\!\cdot\! s & - c\!\cdot\!s & ~~c\!\cdot\!s & - c\!\cdot\!s \\
   ~~c\!\cdot\!s & ~~c\!\cdot\!s & -c\!\cdot\!s & -c\!\cdot\!s \\
 s^2 & -s^2 & -s^2 & ~~s^2 
   \end{pmatrix}  \\
   &=\! \begin{pmatrix} 
   1 & 0 & ~~0 & ~~0 \\
   0& 1\!/\!{\sqrt(2)} &~~1\!/\!{\sqrt(2)} & ~~0 \\
   0& 1\!/\!{\sqrt(2)} & - 1\!/\!{\sqrt(2)} & ~~0 \\
   0 & 0 & ~~0 & -1 
   \end{pmatrix} \!\!
   \begin{pmatrix} 
   2\,c^2 &  ~~0 & ~~0 & ~0\\
   ~0 &  2 \,c \, s & ~~0 & ~0\\
   ~0 &  ~~0    & 2 \,c \, s  & ~0\\
   ~0 &  ~~0    & ~~0 & 2 \, s^2
   \end{pmatrix}  \nonumber \\
 \cdot & \frac{1}{2}  \begin{pmatrix} 
   ~~1 &  ~~1 & ~~1 & ~~1\\
   \sqrt{2} & ~~0 &   ~~0 & -\sqrt{2} \\
   ~~0  & -\sqrt{2} & \sqrt{2} &  ~~0\\
    -1 &  ~~1 & ~~1 & -1
   \end{pmatrix},  \nonumber
\end{align} 
whereas the optimal von Neumann measurement  is:
\begin{align}
&\mathbf{B_2}= \mathbf{U_2} \, \mathbf{V_2}^T =\frac{1}{2}\begin{pmatrix}
   1  & 1 &  1  & 1 \\
   1  & -1 &  1  & -1 \\
   1  & 1 &  -1  & -1 \\
   1  & -1 &  -1  & 1 
   \end{pmatrix}  = \mathbf{B} \otimes \mathbf{B}
\end{align} 
leading to the probability matrix
\begin{equation}
\mathbf{P_2}\!=\! \! \frac{1}{4}\!\! \begin{pmatrix}
   (\!c\!+\!s)^4  & (\!c^2\!-\!s^2)^2 &  (\!c^2\!-\!s^2)^2  & (\!c\!-\!s)^4  \\
   (\!c^2\!-\!s^2)^2  & (\!c\!+\!s)^4  &  (\!c\!-\!s)^4  & (\!c^2\!-\!s^2)^2 \\
   (\!c^2\!-\!s^2)^2  & (\!c\!-\!s)^4  &  (\!c\!+\!s)^4   & (\!c^2\!-\!s^2)^2 \\
   (\!c\!-\!s)^4   & (\!c^2\!-\!s^2)^2 &  (\!c^2\!-\!s^2)^2  & (\!c\!+\!s)^4  
   \end{pmatrix} \!\! = \!\! \mathbf{P} \otimes \mathbf{P}.
\end{equation}

With POVM measurements, the probability of an inconclusive result is derived from the square of the smallest singular value of $\mathbf{\Sigma} $  (i.e., $2 s^2$), leading to $p_{inc} = 1-(2\,s^2)^2$. The average information gain obtained from \eqref{eq:Infgain}, \eqref{eq:Hin_equip} and \eqref{eq:Kfin_POVM} is thus
\begin{equation}
    I_{av} = \log_2(4)-(1-(2\,s^2)^2) \log_2(4)  = 8 \, s^4.
\end{equation}

The quantitative results presented in this section are represented in
Fig.~\ref{fig:avgInfGain} which shows the average information gain for optimal von Neumann and POVM measurements. Individual ($K=1$) and collective ($K=2$) measurements are compared and thus the average information gain is normalized to $K$. The maximum information gain is achieved by the von Neumann measurements, independently from the type of measurement (i.e., individual or collective). The $I_{av}$ advantage of von Neumann with respect to POVM measurements is plotted in Fig.~\ref{fig:diffAvgInfGainC}, which shows that performing collective POVM measurements reduces the information gain. It is possibly to attribute this behavior to the fact that the entropy of inconclusive results is spread over a larger number of signals and increases with the dimension of the spanned Hilbert space.


\section{Conclusions}

The matrix-based formulation of the entropy optimization problem can be optimally solved for non-orthogonal signals, by leveraging on operator algebra and well-known theorems.  We demonstrated that the optimal von Neumann measurements are unique and achieve the same information gain for individual and collective measurements, whether uncorrelated or entangled. By contrast, not only the information gain of POVM is lower than for von Neumann, but also decreasing when increasing the number of signals collectively measured.

The present approach finds direct application to the case of QKD protocols based on weak coherent states.
Moreover, it can be applied to QKD protocol with multiple bases (such as BB84), by computing the average information gain of  each given basis.

\vspace{0.4cm}
\begin{acknowledgments}
The authors wish to thank  Martino Travagnin, Davide Bacco, Claudia De Lazzari and Tobias Osborne for the fruitful discussions on the topic and in particular Adam Lewis for the accurate revision of the manuscript too.

This work was funded by the European Commission as part of the JRC work programme, Project 31920 SatCom4EU.

\end{acknowledgments}

\bibliographystyle{apsrev4-2}
\bibliography{references}
\end{document}


\preprint{APS/123-QED}


\title{\Large{Supplementary Material} \\ \vspace{0.5 cm} 
Can optimal collective measurements outperform
individual measurements for non-orthogonal QKD signals?}

\author{Isabella Cerutti and Petra F. Scudo}

\maketitle

\section{Minimizing the Final entropy}
The problem of minimizing the final entropy, i.e., 
\begin{equation}
   Min \,  H_\mu= Min \, \Big(- \sum_{i=1}^N  Q_{i,\mu} \log(Q_{i,\mu})\Big)
\end{equation}
with the constraint that 
\begin{equation}
   \sum_{i=1}^N Q_{i,\mu} = 1
\end{equation}
can be solved using the method of Lagrange multipliers. 

Let the Lagrange function be defined as
\begin{equation}
    \mathcal{L}(Q_{1,\mu}, Q_{2,\mu}, \dots, \lambda) = -\! \!\sum_i Q_{i,\mu} \log (Q_{i,\mu}) + \lambda \Big( \sum_i Q_{i,\mu} -1\Big).
\end{equation}
The optimization problem can be solved by imposing that the gradient of the Lagrange function is 
\begin{equation}
    \nabla_{Q_{1,\mu}, \dots Q_{2,\mu}, \lambda} \mathcal{L}(Q_{1,\mu}, Q_{2,\mu}, \dots, \lambda) = 0. \label{eq:gradient-null}
\end{equation}
The partial derivatives are
\begin{align}
    \frac{\partial}{\partial Q_{i,\mu}} \mathcal{L}(Q_{1,\mu}, Q_{2,\mu}, \dots, \lambda) \!& = \! - \!\log Q_{i,\mu} - \frac{1}{\ln 2} + \lambda \\
    \frac{\partial}{\partial \lambda} \mathcal{L}(Q_{1,\mu}, Q_{2,\mu}, \dots, \lambda) \!& =  \sum_i  Q_{i,\mu} -1.
\end{align}

By imposing the condition in (\ref{eq:gradient-null}), the following system of equation is obtained
\begin{equation}
\begin{cases}
    - \log Q_{i,\mu} - \frac{1}{\ln 2} +\lambda = 0 & \forall i=1, \dots N\\
     \sum_i  Q_{i,\mu} -1 = 0 &
\end{cases}    
\end{equation}
which has solution 
\begin{equation}
    Q_{i,\mu} = \frac{1}{N}  \quad \forall i=1, \dots N.
\end{equation}

This critical point leads to an entropy of
\begin{equation}
H_\mu=  - \log \frac{1}{N}    
\end{equation}
and it is possible to demonstrate that it is a point of absolute maximum. Since the entropy is a continuous function, the points of absolute minimum are at the border, that is when 
\begin{equation}
    \begin{cases}   
     Q_{j,\mu}  = 1  & j \in {1,2 \dots N}\\
     Q_{i,\mu}  = 0 &  \forall i \neq j, i \in {1,2 \dots N}.
    \end{cases}
\end{equation}